\documentclass[conference]{IEEEtran}
\usepackage{graphicx,amsmath,amssymb,cite,algorithm,algorithmic}
\usepackage{multirow,flushend}
 
\begin{document}
\title{Precoding for Multiuser MIMO Systems with\\Single-Fed Parasitic Antenna Arrays}
\author{
\authorblockN{George~C.~Alexandropoulos, Vlasis~I.~Barousis, and Constantinos~B.~Papadias}
\authorblockA{Broadband Wireless and Sensor Networks Group, Athens Information Technology (AIT)\\ 19.5 km Markopoulo Avenue, 19002 Peania, Athens, Greece}
e-mails: \{alexandg, vbar, cpap\}@ait.gr}
\maketitle

\begin{abstract}
Transmitter (TX) cooperation at various levels has been shown to increase the sum throughput of multiuser multiple-input multiple-output (MIMO) systems. In this paper we consider a $K$-user MIMO system where TXs have only global channel state knowledge. It has been theoretically shown that interference alignment (IA) achieves the $K/2$ degrees of freedom of this $K$-user MIMO interference channel. However, results on IA and all proposed transceiver techniques for this channel up to date, assume conventional antenna arrays at the transceivers with multiple radio-frequency (RF) chains, each connected to a different antenna element. To reduce the consequent hardware burden and power dissipation imposed by such arrays, we propose in this paper the utilization of compact single-RF electronically steerable parasitic (passive) array radiators (ESPARs) at the cooperating TXs. A signal model capable of capturing the characteristics of the considered antenna arrays is first described and then a general precoding design methodology for the tunable parasitic loads at the TXs' ESPARs is introduced. Specific precoding techniques and an indicative ESPAR design are presented for a $3$-user $2\times2$ MIMO system with one ESPAR TX, and the obtained performance evaluation results show that the gains of TX cooperation are still feasible.
\end{abstract}
\thispagestyle{empty}
\IEEEpeerreviewmaketitle

\section{Introduction}\label{sec:Intro}
Interference alignment (IA) is a recently proposed technique for the $K$-user interference channel which is shown to achieve $K/2$ degrees of freedom \cite{J:Jafar_interference}. IA is based on appropriate linear precoding at the transmitters (TXs), aiming at post-receiver processing interference cancellation, and requires only global channel state information at TXs. Exploiting the spatial dimension of multiple-input multiple-output (MIMO) systems, several research works presented precoding designs \cite{J:Jafar_interference, J:Iterative_Jafar, J:Sung_TWC2010, J:Heath_cooperative, J:Luo_TSP_IterMMSE, J:Alexandg_Recon2013} for the $K$-user MIMO interference channel. For the special case of $K=3$, a closed-form solution for IA was presented in \cite{J:Jafar_interference} that was further processed in \cite{J:Sung_TWC2010} for increasing the sum rate. However, for $K>3$ MIMO communicating pairs, closed-form solutions for IA are in general unknown and several iterative algorithms have been recently proposed (see e$.$g$.$ \cite{C:Heath_Globecom, C:Heath_Minimization, J:Sung_TWC2010, J:Iterative_Jafar, J:Heath_cooperative, J:Luo_TSP_IterMMSE, J:Alexandg_Recon2013} and references therein). The vast majority of these algorithms targets at implicitly achieving IA through the optimization of a constrained objective function. To this end, several objective functions have been considered, such as for example: \textit{i}) minimization of the total interference leakage \cite{J:Iterative_Jafar}; \textit{ii}) minimization of the sum of squared errors \cite{C:Heath_Minimization}; \textit{iii}) minimization of the mean squared error \cite{J:Heath_cooperative, J:Luo_TSP_IterMMSE}; \textit{iv}) maximization of the signal-to-interference-plus-noise ratio (SINR)  \cite{J:Iterative_Jafar, J:Heath_cooperative}; and \textit{v}) maximization of the sum-rate performance \cite{J:Sung_TWC2010, J:Alexandg_Recon2013}. 

Commonly in the literature \cite{J:Jafar_interference, C:Heath_Globecom, C:Heath_Minimization, J:Sung_TWC2010, J:Iterative_Jafar, J:Heath_cooperative, J:Luo_TSP_IterMMSE, J:Alexandg_Recon2013}, IA and other transceiver designs for the $K$-user MIMO interference channel assume conventional antenna arrays with uncoupled and uncorrelated elements. Despite their well-known benefits, the significant hardware burden imposed by the use of multiple radio-frequency (RF) chains in such arrays and the consequent increased power consumption requirements discourage their integration especially in lightweight and battery-powered mobile handsets with strict size constraints. A promising adaptive antenna array technology that utilizes a single RF chain to feed a sole antenna element which is surrounded by several parasitic ones was proposed for beamforming by Gyoda and Ohira at the ATR Labs \cite{875370, OhGy00} in $2000$. In the so-called electronically steerable parasitic (passive) array radiators (ESPARs), the feeding at the sole active antenna element induces currents at all adjacent parasitics, thus enabling them to radiate and participate in the shaping of the total radiated beam. In contrast to conventional arrays, ESPAR's radiation is accomplished due to the strong electromagnetic coupling among its elements, which can be achieved in general whenever the inter-element spacing is retained quite low. The strong effective mutual coupling and consequently the currents at the parasitic elements can be further controlled by low-cost and easy-to-implement tunable analog loads attached to them. With ESPARs the TX's front-end hardware complexity is significantly reduced compared with conventional arrays, since easy-to-implement tunable analog loads and as few as one RF chain are used for triggering, instead of the bulkier multiple RF chains. In addition, only one highly linear power amplifier is necessary, which results in significant power savings and higher transmit efficiency. Recently, ESPARs have been proposed for single-RF MIMO systems where spatial multiplexing over the air was demonstrated \cite{HINDAWI_espar_MIMO, ESPAR_MIMO_book, 5669632}. 

Motivated by the recent advances in ESPARs, this paper investigates their potential for precoding for the $K$-user MIMO interference channel. In particular, Section II includes a convenient signal model that is most appropriate for $K$-user MIMO systems with ESPAR TXs. Then, capitalizing on this model in Section III, a general design methodology for arbitrary precoding with ESPARs is introduced. Closed-form expressions and constrained optimization problems for the feeding voltage at the sole active antenna element and the tunable analog loads at the ports of the parasitic antenna elements are derived. In addition, some precoding designs for a $3$-user $2\times2$ MIMO system with an ESPAR-equipped TX are presented. Finally, Section IV contains indicative performance evaluation results for this system with a realistic ESPAR design, which are also compared against the ones obtained with an ideal conventional antenna array. 

\section{Signal, System and Channel Model}\label{sec:IAespar}
In the following, we first describe the considered signal model that captures the functionality of a TX equipped with an arbitrary array. In addition, we present the considered $K$-user MIMO system and channel model.

\subsection{Signal Model}\label{sec:Sig}
Assume a TX $k$ equipped with a conventional antenna array consisting of $n_{\rm T}^{[k]}$ elements arbitrarily arranged in space. The $i$-th antenna element, with $i=1,2,\ldots,n_{\rm T}^{[k]}$, is fed by an input source with voltage $V_i^{[k]}$ and output impedance $Z_i^{[k]}$, and in the most general case the arbitrary antenna placement results in the generation of a mutual coupling matrix $\mathbf{Z}_{\rm T}^{[k]}\in\mathbb{C}^{n_{\rm T}^{[k]}\times n_{\rm T}^{[k]}}$. Then, the complex-valued vector with the $n_{\rm T}^{[k]}$ currents at the ports of the array of TX $k$ is given by the generalized Ohm's law as
\begin{equation}\label{Eq:currents_arbitrary_array}
\mathbf{i}_k = \left(\mathbf{Z}_{\rm T}^{[k]}+\mathbf{Z}_{\rm G}^{[k]}\right)^{-1}\mathbf{v}_k\triangleq\mathbf{D}_{\rm T}^{[k]}\mathbf{v}_k
\end{equation}   
where $\mathbf{Z}_{\rm G}^{[k]}\in\mathbb{C}^{n_{\rm T}^{[k]}\times n_{\rm T}^{[k]}}$ is a diagonal matrix with the output impedances $Z_i^{[k]}$'s in the main diagonal, $\mathbf{v}_k\in\mathbb{C}^{n_{\rm T}^{[k]}\times1}$ is the vector with the ports' feeding voltages and $\mathbf{D}_{\rm T}^{[k]}\in\mathbb{C}^{n_{\rm T}^{[k]}\times n_{\rm T}^{[k]}}$ represents the effective mutual coupling matrix. 

In the special case where TX $k$ is equipped with an ideal array, i$.$e$.$ uncoupled and uncorrelated antenna elements, the coupling matrix degenerates to $\mathbf{Z}_{\rm T}^{[k]} = Z_c^{[k]}\mathbf{I}_{n_{\rm T}^{[k]}}$, with $Z_c^{[k]}$ denoting the common self impedance of the diverse elements that is assumed to match adequately the source, where $\mathbf{I}_{n_{\rm T}^{[k]}}$ is the $n_{\rm T}^{[k]}\times n_{\rm T}^{[k]}$ identity matrix. Furthermore, $\mathbf{Z}_{\rm G}^{[k]}= R_0^{[k]}\mathbf{I}_{n_{\rm T}^{[k]}}$, where $R_0^{[k]}=50$ $\Omega$ is a typical value for the output resistance of a source. Therefore, $\mathbf{D}_{\rm T}^{[k]}$ in \eqref{Eq:currents_arbitrary_array} simplifies to $\mathbf{D}_{\rm T}^{[k]}=(Z_c^{[k]}+R_0^{[k]})\mathbf{I}_{n_{\rm T}^{[k]}}$ and hence $\mathbf{i}_k=(Z_c^{[k]}+R_0^{[k]})^{-1}\mathbf{v}_k$, i$.$e$.$ the ports' currents are just a scaled version of the ports' feeding voltages.
		
\begin{figure}[!t]
\centering
\includegraphics[keepaspectratio,width=3.5in]{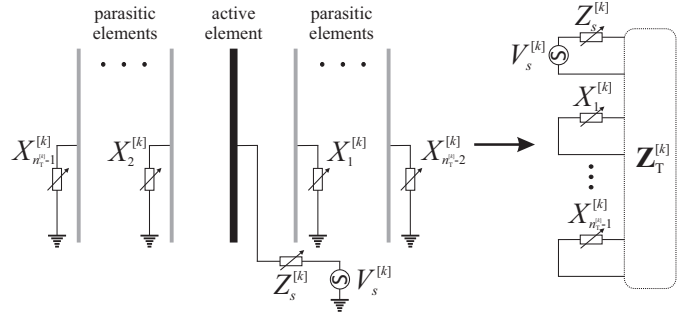}
\caption{Single-fed ESPAR at TX $k$ and the equivalent antenna model.}
\label{Fig:ESPAR}
\end{figure}

In the case of a TX $k$ equipped with a single-fed ESPAR as shown in Fig.~\ref{Fig:ESPAR}, only the sole active antenna element is fed by a voltage $V_s^{[k]}$, while the remaining $n_{\rm T}^{[k]}-1$ elements are excited passively due to strong mutual coupling. The antenna couplings, and in turn the currents at the ports of the parasitic elements, can be controlled by tuning appropriately the $n_{\rm T}^{[k]}-1$ analog loads $X_n^{[k]}$'s $\in\mathbb{C}$ (with $n=1,2,\ldots,n_{\rm T}^{[k]}-1$) attached to the parasitics. Hence, the vector with the currents at the $n_{\rm T}^{[k]}$ elements of ESPAR TX $k$ is obtained similar to \eqref{Eq:currents_arbitrary_array} as
\begin{equation}\label{Eq:currents_ESPAR_array}
\mathbf{i}_k = \left(\mathbf{Z}_{\rm T}^{[k]}+\mathbf{X}_k\right)^{-1}[V_s^{[k]}\,\,\underbrace{0\,\,\ldots\,\,0}_{n_{\rm T}^{[k]}-1}]^{\rm T}.
\end{equation}
In \eqref{Eq:currents_ESPAR_array}, $\mathbf{X}_k={\rm diag}([Z_s^{[k]}\,\,\bar{\mathbf{x}}_k^{\rm T}])\in\mathbb{C}^{n_{\rm T}^{[k]}\times n_{\rm T}^{[k]}}$ is a diagonal matrix that contains the tunable output resistance $Z_s^{[k]}$ of the sole active element, which is used for dynamic matching. Moreover, $\mathbf{X}_k$ includes $\bar{\mathbf{x}}_k\in\mathbb{C}^{(n_{\rm T}^{[k]}-1)\times1}$ having the $X_n^{[k]}$'s. 

Using \eqref{Eq:currents_arbitrary_array} and \eqref{Eq:currents_ESPAR_array} for $\mathbf{i}_k$ of TX $k$ with conventional arrays and single-fed ESPARs, respectively, the open-circuit voltage vector $\mathbf{y}\in\mathbb{C}^{n_{\rm R}^{[k]}\times 1}$, with the voltages due to the impinging signal at the ports of a $n_{\rm R}^{[k]}$-element receiver (RX) $k$, can be expressed as
\begin{equation}\label{Eq:signal_model}
\mathbf{y}_k = \mathbf{H}_{k,k}\mathbf{i}_k + \mathbf{n}_k.
\end{equation}  
In \eqref{Eq:signal_model}, $\mathbf{H}_{k,k}\in\mathbb{C}^{n_{\rm R}^{[k]}\times n_{\rm T}^{[k]}}$ is the channel matrix whose entries relate the input currents at TX $k$ with the output open-circuit voltages at RX $k$, and $\mathbf{n}_k\in\mathbb{C}^{n_{\rm R}^{[k]}\times 1}$ denotes the zero-mean complex additive white Gaussian noise (AWGN) vector with covariance matrix $\sigma^2_k\mathbf{I}_{n_{\rm R}^{[k]}}$.

\subsection{System and Channel Model}\label{sec:Sys}
Suppose a multiuser MIMO system consisting of $K$ pairs of communicating users. In particular, each TX $k$, where $k=1,2,\ldots,K$, equipped with $n_{\rm T}^{[k]}$ antennas, wishes to communicate with its intended $n_{\rm R}^{[k]}$-antenna RX $k$. All $K$ simultaneous transmissions of symbols $\mathbf{s}_k\in\mathbb{C}^{d_k\times 1}$, with $d_k\leq\min(n_{\rm T}^{[k]}, n_{\rm R}^{[k]})$ $\forall\,k$, are assumed perfectly synchronized and each TX $k$ processes individually its $\mathbf{s}_k$ with a linear matrix $\mathbf{F}_k\in\mathbb{C}^{n_{\rm T}^{[k]}\times d_k}$ before transmission. For the transmitted power per TX $k$ it is assumed that $\mathbb{E}\{||\mathbf{F}_k\mathbf{s}_k||^2\}\leq {\rm P}$ with ${\rm P}$ being the total power constraint per TX and $\mathbb{E}\{\cdot\}$ denoting expectation. Using the signal model in \eqref{Eq:signal_model} and by setting $\mathbf{i}_k=\mathbf{F}_k\mathbf{s}_k$ $\forall\,k$, the open-circuit voltage vector at each RX $k$ can be mathematically expressed as
\begin{equation}\label{Eq:System}
\mathbf{y}_k = \mathbf{\hat{H}}_{k,k}\mathbf{F}_k\mathbf{s}_k + \sum_{\ell=1,\ell\neq k}^K\mathbf{\hat{H}}_{k,\ell}\mathbf{F}_\ell\mathbf{s}_\ell + \mathbf{n}_k
\end{equation}
where $\mathbf{\hat{H}}_{k,\ell}\in\mathbb{C}^{n_{\rm R}^{[k]}\times n_{\rm T}^{[\ell]}}$, with $k,\ell=1,2,\ldots,K$, denotes the channel matrix between RX $k$ and TX $\ell$.

We assume TXs equipped with either ideal uncoupled and uncorrelated uniform linear arrays (ULAs) or single-fed ESPARs whereas, RXs are assumed to have ideal uncoupled and uncorrelated ULAs. Depending on the type of the considered array at TX $\ell$, $\mathbf{\hat{H}}_{k,\ell}$ is modeled as
\begin{equation}\label{Eq:channel_selection}
\mathbf{\hat{H}}_{k,\ell} =
\left\{
\begin{array}{lr}
\mathbf{H}_{k,\ell},&{\rm ULA}\\
\mathbf{H}_{k,\ell}\left(\mathbf{R}_{\ell}^{1/2}\right)^{\rm T},&{\rm ESPAR}
\end{array}
\right.
\end{equation}
where the elements of $\mathbf{H}_{k,\ell}$ are independent and identically distributed as circularly-symmetric complex Gaussian random variables with zero mean and unit variance. In addition, for the case of an ESPAR TX where the antenna inter-element spacing is required to be quite low \cite{ESPAR_MIMO_book}, $\mathbf{R}_{\ell}\in\mathbb{C}^{n_{\rm T}^{[\ell]}\times n_{\rm T}^{[\ell]}}$ represents the transmit-side spatial correlation matrix. It is noted that, although in non-ideal ULAs any couplings among its antenna elements are in general an intrinsic component of the channel model $\mathbf{H}_{k,\ell}$, in the ESPAR channel modeling in the second branch of \eqref{Eq:channel_selection} they are not. The effective mutual coupling in ESPARs, i$.$e$.$ $\mathbf{D}_{\rm T}^{[k]}=(\mathbf{Z}_{\rm T}^{[k]}+\mathbf{X}_k)^{-1}$ in \eqref{Eq:currents_ESPAR_array}, is a design parameter that is used to adjust the ports' currents at the parasitic elements to the desired array functionality; this is accomplished by tuning appropriately the analog loadings $X_n^{[k]}$'s connected to them. 

\section{ESPAR Precoding for Multiuser MIMO}\label{sec:CfIA_with_ESPAR}
We first introduce a general precoding methology for $K$-user MIMO systems with single-fed ESPAR TXs and then present two specific precoding designs for a $3$-user $2\times2$ MIMO system consisting of two TXs with ideal uncoupled and uncorrelated ULAs and one single-fed ESPAR TX.

\subsection{Precoding Design Methodology}\label{sec:Method}
As described in Section~\ref{sec:Sys}, to achieve the desired precoding vector $\mathbf{F}_k\mathbf{s}_k$ at TX $k$, its antenna ports' currents need to be designed as $\mathbf{i}_k=\mathbf{F}_k\mathbf{s}_k$. By substituting \eqref{Eq:currents_arbitrary_array} in the latter expression yields
\begin{equation}\label{Eq:General_Equation}
\left(\mathbf{Z}_{\rm T}^{[k]}+\mathbf{Z}_{\rm G}^{[k]}\right)^{-1}\mathbf{v}_k = \mathbf{F}_k\mathbf{s}_k.
\end{equation}
When a TX $k$ is equipped with an ideal ULA, the desired currents at its antenna ports are controlled externally by $n_{\rm T}^{[k]}$ distinct input sources with feeding voltages that are easily derived using Section~\ref{sec:Sig} and \eqref{Eq:General_Equation} as $\mathbf{v}_k=(Z_c^{[k]}+R_0^{[k]})\mathbf{F}_k\mathbf{s}_k$. In the case of a ESPAR TX $k$, replacing \eqref{Eq:currents_ESPAR_array} in \eqref{Eq:General_Equation} results in the following expression 
\begin{equation}\label{Eq:ESPAR_Equation}
\left(\mathbf{Z}_{\rm T}^{[k]}+\mathbf{X}_k\right)^{-1}[V_s^{[k]}\,\,0\,\,\ldots\,\,0]^{\rm T} = \mathbf{F}_k\mathbf{s}_k
\end{equation}
from which $\mathbf{X}_k$ and $V_s^{[k]}$ need to be computed. 

Using the results of \cite{WCL_espar_cf} to solve \eqref{Eq:ESPAR_Equation}, $V_s^{[k]}$ and $\bar{\mathbf{x}}_k$ at a single-fed ESPAR TX $k$ can be obtained in closed-form as
\begin{subequations}\label{Eq:ESPAR_cf}
\begin{equation}\label{Eq:Voltage}
V_s^{[k]} = \left[\mathbf{Z}_{\rm T}^{[k]}+\mathbf{X}_k\right]_{1,:}\mathbf{F}_k\mathbf{s}_k
\end{equation}
\begin{equation}\label{Eq:Loadings}
X_n^{[k]} = -\left(\left[\mathbf{F}_k\mathbf{s}_k\right]_{n}\right)^{-1}\left[\mathbf{Z}_{\rm T}^{[k]}\right]_{n+1,:}\mathbf{F}_k\mathbf{s}_k
\end{equation}
\end{subequations}
where $n=1,2,\ldots,n_{\rm T}^{[k]}-1$, and notations $[\mathbf{A}]_{n,:}$ and $[\mathbf{a}]_{n}$ represent the $n$-th row of matrix $\mathbf{A}$ and the $n$-th element of vector $\mathbf{a}$, respectively. However, to ensure that the ESPAR can support the desired precoding vector $\mathbf{F}_k\mathbf{s}_k$, the following condition needs to hold \cite{WCL_espar_cf}
\begin{equation}\label{Eq:ESPAR_Constraint}
{\rm Re}\left\{\left(\left[\mathbf{F}_k\mathbf{s}_k\right]_{1}\right)^{-1}\left[\mathbf{Z}_{\rm T}^{[k]}\right]_{1,:}\mathbf{F}_k\mathbf{s}_k\right\}>0.
\end{equation}
Furthermore, the input impedance in the ESPAR's single feeding port is a function of the parasitic analog loadings \cite{Balanis}, and it can be shown that is given by
\begin{equation}\label{Eq:Input_Impedance}
\begin{split}
&Z_{\rm in}^{[k]}(\bar{\mathbf{x}}_k) = \left[\mathbf{Z}_{\rm T}^{[k]}\right]_{1,1} - \left[\mathbf{Z}_{\rm T}^{[k]}\right]_{1,2:n_{\rm T}^{[k]}}
\\&\times\left\{\left[\mathbf{Z}_{\rm T}^{[k]}\right]_{2:n_{\rm T}^{[k]},2:n_{\rm T}^{[k]}}+{\rm diag}(\bar{\mathbf{x}}_k)\right\}^{-1}\left[\mathbf{Z}_{\rm T}^{[k]}\right]_{2:n_{\rm T}^{[k]},1}
\end{split}
\end{equation}
where $\left[\mathbf{A}\right]_{k,\ell}$ is the $(k,\ell)$-th element of matrix $\mathbf{A}$ whereas, notation $[\mathbf{A}]_{m:n,k:\ell}$ represents the submatrix of $\mathbf{A}$ obtained from its $m,m+1,\ldots,n-1,n$ rows and $k,k+1,\ldots,\ell-1,\ell$ columns. Thus, different $\bar{\mathbf{x}}_k$ causes different mismatch effects between the ESPAR and the feeding source, which may degrade the array's performance. To guarantee negligible return losses due to mismatch, $Z_{\rm in}^{[k]}(\bar{\mathbf{x}}_k)$ needs to be dynamically matched by $Z_s^{[k]}$. Finally, the values of $\bar{\mathbf{x}}_k$ and $Z_s^{[k]}$ are usually restricted in practice to certain ranges of values.

To handle cases where \eqref{Eq:ESPAR_Constraint} and/or any of the aforementioned requirements for $\bar{\mathbf{x}}_k$ and $Z_s^{[k]}$ is not fulfilled, constraint optimization approaches can be adopted for computing $V_s^{[k]}$ and $\mathbf{X}_k$, as in \cite{Fast_BF}. In this paper, using the formulation described by \eqref{Eq:ESPAR_Equation} for precoding at a single-fed ESPAR TX $k$, the constrained minimization of the following objective function is considered
\begin{equation}\label{Eq:cost_func_general}
f_k\left(\mathbf{X}_k,V_s^{[k]}\right) = \left\|\mathbf{F}_k\mathbf{s}_k-\left(\mathbf{Z}_{\rm T}^{[k]}+\mathbf{X}_k\right)^{-1}[V_s^{[k]}\,\,0\,\,\ldots\,\,0]^{\rm T}\right\|_{\rm F}^2
\end{equation}
where $||\cdot||_{\rm F}$ denotes the Frobenius norm.

For the special case where $d_k=1$ (single-stream transmission) and after setting $\mathbf{s}_k$ and $\mathbf{F}_k$ equal to $s_k\in\mathbb{C}$ and $\mathbf{f}_k\in\mathbb{C}^{n_{\rm T}^{[k]}\times1}$, respectively, in \eqref{Eq:ESPAR_cf}, it can be easily seen that \eqref{Eq:Voltage} simplifies to $V_s^{[k]} = [\mathbf{Z}_{\rm T}^{[k]}+\mathbf{X}_k]_{1,:}\mathbf{f}_k s_k$, \eqref{Eq:Loadings} to $X_n^{[k]} = -([\mathbf{f}_k]_{n})^{-1}[\mathbf{Z}_{\rm T}^{[k]}]_{n+1,:}\mathbf{f}_k$ and similarly does \eqref{Eq:ESPAR_Constraint}. The latter expressions indicate that for this special case, $s_k$ feeds only the sole active antenna element and does not determine the values of $\bar{\mathbf{x}}_k$. However, replacing the latter expressions in \eqref{Eq:cost_func_general} results in an expression for $f_k(\mathbf{X}_k,V_s^{[k]})$ that includes $s_k$. This means that whenever the constrained minimization of $f_k(\mathbf{X}_k,V_s^{[k]})$ is necessary, this needs to be done on a per symbol basis. An alternative approach that alleviates this need can be summarized as follows. By replacing $\mathbf{F}_k$ with $\mathbf{f}_k$ and $\mathbf{s}_k$ with $s_k$ in \eqref{Eq:ESPAR_Equation} and after some basic algebraic manipulations, it can be easily shown that
\begin{equation}\label{Eq:ESPAR_Equation_manipulated}
V_s^{[k]}\mathbf{b}_k\left(\mathbf{X}_k\right) = s_k\mathbf{f}_k
\end{equation}
where $\mathbf{b}_k(\mathbf{X}_k)=[(\mathbf{Z}_{\rm T}^{[k]}+\mathbf{X}_k)^{-1}]_{:,1}\in\mathbb{C}^{n_{\rm T}^{[k]}\times1}$ with notation $[\mathbf{A}]_{:,n}$ representing the $n$-th column of matrix $\mathbf{A}$. Hence, after solving \eqref{Eq:ESPAR_Equation_manipulated}, yields $V_s^{[k]}=s_k$ and $\mathbf{b}_k(\mathbf{X}_k) = \mathbf{f}_k$. The former expression implies that the voltage of the sole active antenna element needs to be set proportionally to $s_k$ and the latter indicates that the diagonal elements of $\mathbf{X}_k$ do not depend on $s_k$, but only on the desired precoding vector $\mathbf{f}_k$ and the mutual coupling matrix $\mathbf{Z}_{\rm T}^{[k]}$. Closed-form expressions for $\bar{\mathbf{x}}_k$ as well as necessary conditions for the desired ESPAR functionality can be obtained for certain cases similar to \eqref{Eq:Loadings} and \eqref{Eq:ESPAR_Constraint}, respectively, but are omitted here for brevity. Furthermore, the objective function $f_k(\mathbf{X}_k,V_s^{[k]})$ in \eqref{Eq:cost_func_general} can be modified as
\begin{equation}\label{Eq:cost_func_d1}
f_k\left(\mathbf{X}_k\right) = \left\|\mathbf{f}_k-\mathbf{b}_k\left(\mathbf{X}_k\right)\right\|^2
\end{equation}
which is now independent of $s_k$ and is determined only by $\mathbf{f}_k$ and $\mathbf{Z}_{\rm T}^{[k]}$.

\subsection{A $3$-User $2\times2$ MIMO Example}\label{sec:Example}
We consider $3$-user MIMO system where TX $2$ is equipped with a $2$-element single-fed ESPAR whereas, all other users have ideal ULAs each with $2$ antenna elements. It is also assumed that RXs estimate perfectly their intended and unintended channels, and TXs have global channel state know\-ledge. The optimum sum-rate multiplexing gain of this $3$-user $2\times2$ MIMO interference channel is $3$ and is achieved by IA \cite{J:Jafar_interference}. We therefore set the IA feasibility conditions as $d_k=1$ $\forall\,k=1,2$ and $3$ \cite{J:Yetis}, and summarize two representative precoding designs for $\mathbf{f}_k$'s based on the models described in Section~\ref{sec:IAespar}.  
\subsubsection{Closed-Form IA}\label{sec:CfIA}
Replacing $\mathbf{F}_k$'s with $\mathbf{f}_k$'s and $\mathbf{s}_k$'s with $s_k$'s in \eqref{Eq:System} and using \eqref{Eq:channel_selection}, the IA-achieving precoding vectors at TXs $1$, $2$ and $3$, respectively, are given by \cite{J:Jafar_interference}
\begin{subequations}\label{Eq:V_IA}
\begin{equation}\label{Eq:V_IA_1}
\mathbf{f}_1 = v_{\rm max}\left(\mathbf{H}_{3,1}^{-1}\mathbf{H}_{3,2}\mathbf{H}_{1,2}^{-1}\mathbf{H}_{1,3}\mathbf{H}_{2,3}^{-1}\mathbf{H}_{2,1}\right)
\end{equation}
\begin{equation}\label{Eq:V_IA_2}
\mathbf{f}_2 = \left(\mathbf{R}_{2}^{1/2}\right)^{\rm -T}\mathbf{H}_{3,2}^{-1}\mathbf{H}_{3,1}\mathbf{f}_1
\end{equation}
\begin{equation}\label{Eq:V_IA_3}
\mathbf{f}_3 = \mathbf{H}_{2,3}^{-1}\mathbf{H}_{2,1}\mathbf{f}_1
\end{equation}
\end{subequations}
where $v_{\rm max}(\mathbf{A})$ represents the eigenvector of matrix $\mathbf{A}$ corresponding to its largest eigenvalue and $\mathbf{A}^{\rm -T}$ denotes the inverse of the transpose of $\mathbf{A}$. 

\subsubsection{Maximum SINR}\label{sec:MaxSINR}
This algorithm aims at maximizing the SINR of $s_k$ $\forall\,k=1,2$ and $3$, and outperforms closed-form IA when noise is the dominant degradation factor in the multiuser channel. It capitalizes on the reciprocity of time division duplexing channels and adopts an alternating optimization approach to compute $\mathbf{f}_k$'s and each receive vector $\mathbf{u}_k\in\mathbb{C}^{2\times1}$ at RX $k$. If the iterative approach converges to a set of $\mathbf{f}_k$'s and $\mathbf{u}_k$'s or the number of maximum algorithmic iterations is reached, the unit-norm precoding vector at each TX $k$ is given by \cite{J:Iterative_Jafar}
\begin{equation}\label{Eq:V_IA_3}
\mathbf{f}_k = \frac{\mathbf{C}_k^{-1}\mathbf{\hat{H}}_{k,k}^{\rm H}\mathbf{u}_k}{\left\|\mathbf{C}_k^{-1}\mathbf{\hat{H}}_{k,k}^{\rm H}\mathbf{u}_k\right\|}
\end{equation}
where $||\cdot||$ denotes the Euclidean norm. In \eqref{Eq:V_IA_3}, $\mathbf{C}_k\in\mathbb{C}^{2\times2}$ comprises of the covariance matrices of the noise and inter-user interference, and is obtained as
\begin{equation}\label{Eq:C_maxSINR}
\mathbf{C}_k = {\rm P}\sum_{\ell=1,\ell\neq k}^{3}\mathbf{\hat{H}}_{\ell,k}^{\rm H}\mathbf{u}_\ell\mathbf{u}_\ell^{\rm H}
\mathbf{\hat{H}}_{\ell,k} + \sigma^2_k\mathbf{I}_{2}.
\end{equation}
Furthermore, the unit-norm $\mathbf{u}_k$'s needed for the computation of \eqref{Eq:V_IA_3} are given by similar expressions to $\mathbf{f}_k$'s \cite{J:Iterative_Jafar}.

\subsubsection{Achievable Performance}\label{sec:SumRate}
In general it is very difficult to analyze the maximum performance of MIMO transmission with ESPAR precoding under realistic constraints on ESPAR design and transmit power. As shown in \eqref{Eq:cost_func_general} for $d_k>1$, $\mathbf{X}_k$ depends both on $\mathbf{F}_k$ and $\mathbf{s}_k$, and assessment of the mutual information between the actually transmitted and received signal is needed. For the considered $3$-user $2\times2$ MIMO system though, it can be easily seen from \eqref{Eq:cost_func_d1} that $\mathbf{X}_2$ in the ESPAR TX $2$ depends only on $\mathbf{f}_2$ and not on $s_2$. In addition, it can be shown that the mutual information of this system is maximized under an average transmit power constraint when circularly symmetric complex Gaussian signalling $s_2$ is used. Hence, to investigate the perfomance of the considered multiuser precoding designs we have obtained the ergodic sum-rate performance, which is mathematically obtained as
\begin{equation}\label{Eq:Erg_SumRate}
\mathcal{R} = \mathbb{E}_{\mathbf{\hat{H}}}\left\{\sum_{k=1}^3\log_2\left[\det\left(\mathbf{I}_2+{\rm P}\mathbf{\hat{H}}_{k,k}\mathbf{f}_k\mathbf{f}_k^{\rm H}\mathbf{\hat{H}}_{k,k}^{\rm H}\mathbf{Q}_k^{-1}\right)\right]\right\}
\end{equation}
where $\mathbb{E}_{\mathbf{\hat{H}}}\{\cdot\}$ denotes the expectation ovel all channel reali\-zations $\mathbf{\hat{H}}_{k,\ell}$ $\forall\,k,\ell=1,2$ and $3$, and $\mathbf{Q}_k\in\mathbb{C}^{2\times2}$ is the interference-plus-noise covariance matrix at each RX $k$, which is given by
\begin{equation}\label{Eq:IpN_Cov}
\mathbf{Q}_k = {\rm P}\sum_{\ell=1,\ell\neq k}^{3}\mathbf{\hat{H}}_{k,\ell}\mathbf{f}_\ell\mathbf{f}_\ell^{\rm H}
\mathbf{\hat{H}}_{k,\ell}^{\rm H} + \sigma^2_k\mathbf{I}_{2}.
\end{equation}
As shown in Section~\ref{sec:Method}, the desired precoding vector at ESPAR TX $2$, i$.$e$.$ $\mathbf{f}_2$, is designed as $\mathbf{b}_2(\mathbf{X}_2)$ with $\mathbf{X}_2={\rm diag}([Z_s^{[k]}\,\,X_1^{[2]}])$, which is obtained in the most general case from the constrained minimization of \eqref{Eq:cost_func_d1} given $\mathbf{f}_2$. Note, however, that this approach may often result in the occurrence of an error vector $\mathbf{e}_2\in\mathbb{C}^{2\times1}$ such that $\mathbf{b}_2(\mathbf{X}_2)=\mathbf{f}_2+\mathbf{e}_2$.  

\section{Performance Evaluation Results}\label{sec:Results}
In this section we evaluate the performance of the $3$-user $2\times2$ MIMO system presented in Section~\ref{sec:Example}. A realistic ESPAR with $2$ dipole elements, one active and one parasitic, that resonates at the center frequency $f_c=2.4$ GHz has been considered for TX $2$ and designed in the IE3D electromagnetic software package. The ESPAR's elements are printed on a FR4 substrate of height $0.8$ mm with dielectric constant $\epsilon_r=4.45$ and dielectric loss tangent $\tan\delta=0.017$. Each element is $1.4$ mm of width and $47$ mm of height, while the inter-element spacing is $0.14\lambda$ with $\lambda$ being the wavelength. The coupling matrix $\mathbf{Z}_{\rm T}^{[2]}$ of the designed ESPAR is obtained from IE3D in order to capture the effects of the structural characteristics of the ESPAR, such as the element dimensions, substrate, inter-element spacing and other electromagnetic effects. 
\begin{figure}[!t]
\centering{\includegraphics[height=2.8in,width=2.8in]{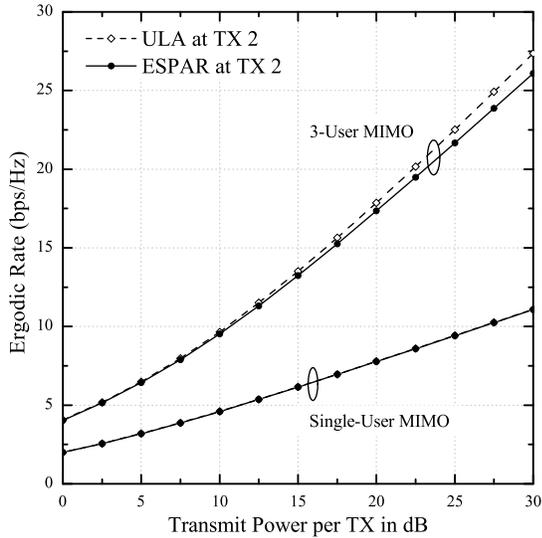}}
\caption{Ergodic sum-rate performance, $\mathcal{R}$, versus transmit power per Tx, ${\rm P}$, for the considered $3$-user $2\times2$ MIMO system with the closed-form IA precoding.}
\label{Fig:cfIA}
\end{figure}

Complex values have been assumed for $X_1^{[2]}$, capitilizing on the switched complex loading circuit recently presented in \cite{HINDAWI_espar_MIMO}. For the cases where the value for $X_1^{[2]}$, obtained from the solution of $\mathbf{b}_2(\mathbf{X}_2) = \mathbf{f}_2$, does not meet realistic constraints, we utilized a genetic algorithm for minimizing $f_2\left(\mathbf{X}_2\right)$ in \eqref{Eq:cost_func_d1}. The considered constraints are the following: \textit{i}) Both the real and the imaginary parts of $X_1^{[2]}$ and $Z_s^{[2]}$ were restricted between $-500$ and $500$ $\Omega$; and \textit{ii}) The return losses were constrained to be less than $-10$ dB at the resonant frequency of the ESPAR. It is noted that the return losses in dB are obtained as $r = 10\log_{10}\left(|\rho|^2\right)$ with $\rho$ given by
\begin{equation}\label{Eq:return_losses}
\rho = \frac{Z_{\rm in}^{[2]}\left(X_1^{[2]}\right)-Z_s^{[2]*}}{Z_{\rm in}^{[2]}\left(X_1^{[2]}\right)+Z_s^{[2]}}
\end{equation}
where $Z_s^{[2]*}$ is the complex conjugate of $Z_s^{[2]}$.

\begin{figure}[!t]
\centering{\includegraphics[height=2.8in,width=2.8in]{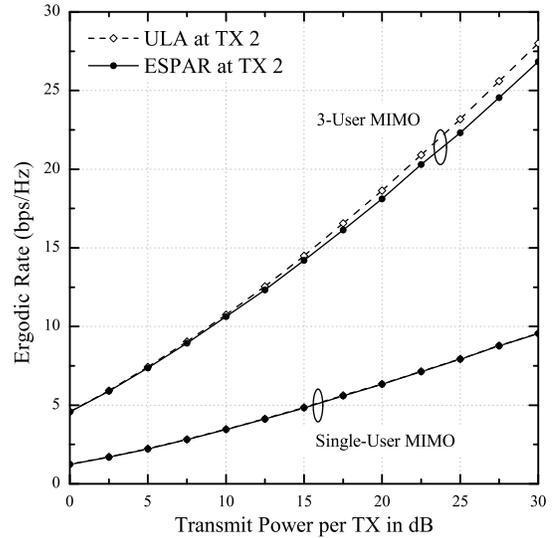}}
\caption{Ergodic sum-rate performance, $\mathcal{R}$, versus transmit power per Tx, ${\rm P}$, for the considered $3$-user $2\times2$ MIMO system with the maximum SINR precoding.}
\label{Fig:MaxSINR}
\end{figure}

The ergodic sum-rate performance $\mathcal{R}$ of the considered $3$-user $2\times2$ MIMO system with the closed-form IA and maximum SINR precoding schemes is depicted in Figs.~\ref{Fig:cfIA} and \ref{Fig:MaxSINR}, respectively, as a function of the transmit power ${\rm P}$ of every TX for fixed noise variance $\sigma_k^2=1$ at each RX $k$. The results were obtained after averaging over $1000$ random channel realizations. Within these figures, the ergodic rate of the single MIMO link between TX $2$ and RX $2$ is also illustrated for the same precoding techniques as those considered for obtaining $\mathcal{R}$. As clearly seen from both figures and as expected, when ${\rm P}$ increases, the performance of single-link and $3$-user MIMO improves independently of the considered precoding technique and the employed array at TX $2$. Furthermore, by comparing $\mathcal{R}$ in Figs.~\ref{Fig:cfIA} and \ref{Fig:MaxSINR}, it is obvious that, when interference is weak, the maximum SINR precoding outperforms closed-form IA whereas, in the interference-limited regime, both precoding techniques have similar performance. The latter expected behavior \cite{J:Iterative_Jafar} happens both for a ULA-equipped TX $2$ and when this TX employs our designed single-fed ESPAR. In addition, it is shown in both figures that for all considered power levels, the ergodic rate of the single MIMO link is almost the same when TX $2$ is equipped with an ideal uncoupled and uncorrelated ULA and when it is equipped with the designed ESPAR. This shows that the proposed precoding design methodology is very accurate and whenever a non-zero $\mathbf{e}_2$ occurs it is very low. The aforementioned trend holds also for $\mathcal{R}$ when ${\rm P}<15$ dB. However, for higher values of ${\rm P}$, $\mathcal{R}$ with the designed ESPAR is slightly lower than that with the ULA, and as a result the sum-rate mutiplexing gain is less than the maximum value of $3$. This happens due to the fact that there exists some negligible error $\mathbf{e}_2$ that does allow to align interference at RXs for a certain ${\rm P}$ value and above.

\section{Conclusion}\label{sec:Conclusion}
In this paper, precoding for multiuser MIMO systems with single-RF ESPAR TXs has been considered. By introducing a signal model that captures the effect of TXs equipped with ESPARs, a general design methodology for multiuser MIMO precoding was presented. In particular, we have provided closed-form expressions and constrained optimization pro\-blems for the feeding voltage at the sole active antenna element and the tunable analog loads at the ports of the parasitic ones. Moreoever, precoding designs with closed-form IA and the maximum SINR algorithm were presented for a $3$-user $2\times2$ MIMO system with one single-RF ESPAR TX. Results with a realistic but common ESPAR design were obtained and showed that the ergodic sum rate of the considered system with the ESPAR TX is very close to that with an ideal uncoupled and uncorrelated ULA. We expect that a more advanced antenna design, optimized for the considered multiuser system, is feasible and will lead to improved results in the interference-limited regime.

\section*{Acknowledgment}
This work has been supported by EU FP7 Specific Targeted Research Project HARP under grant number 318489. 

\bibliographystyle{IEEEtran}
\bibliography{IEEEabrv,refs}

\end{document}